\documentclass[runningheads]{llncs}
\usepackage[T1]{fontenc}
\usepackage{graphicx}

\RequirePackage{jabbrv}
\usepackage[colorlinks, linkcolor = blue, citecolor = blue, filecolor = blue,
urlcolor = blue]{hyperref}
\usepackage{url}
\usepackage{amsmath}
\usepackage{amssymb}
\usepackage{mathtools}
\usepackage{geometry}
\usepackage{subcaption}
\usepackage[biblabel]{cite}
\renewcommand{\url}[1]{}
\renewcommand{\month}[1]{}
\begin{document}
\title{Benchmarking Data Efficiency in \texorpdfstring{$\Delta$}{Delta}-ML and Multifidelity Models for Quantum Chemistry}
\titlerunning{MFML Data Efficiency}

\author{Vivin Vinod\inst{1}\orcidID{0000-0001-6218-5053} \and
Peter Zaspel\inst{1}\orcidID{0000-0002-7028-6580}}
\authorrunning{V. Vinod \& P. Zaspel}

\institute{
School of Mathematics and Natural Sciences, University of Wuppertal, Wuppertal 42119, Germany\\
\email{\{vinod,zaspel\}@uni-wuppertal.de}
}
\maketitle

\begin{abstract}
The development of machine learning (ML) methods has made quantum chemistry (QC) calculations more accessible by reducing the compute cost incurred in conventional QC methods. This has since been translated into the overhead cost of generating training data. Increased work in reducing the cost of generating training data resulted in the development of $\Delta$-ML and multifidelity machine learning methods which use data at more than one QC level of accuracy, or fidelity. 

This work compares the data costs associated with $\Delta$-ML, multifidelity machine learning (MFML), and optimized MFML (o-MFML) in contrast with a newly introduced Multifidelity$\Delta$-Machine Learning (MF$\Delta$ML) method for the prediction of ground state energies, vertical excitation energies, and the magnitude of electronic contribution of molecular dipole moments from the multifidelity benchmark dataset QeMFi. This assessment is made on the basis of training data generation cost associated with each model and is compared with the single fidelity kernel ridge regression (KRR) case. 
The results indicate that the use of multifidelity methods surpasses the standard $\Delta$-ML approaches in cases of a large number of predictions. 
For applications which require only a few evaluations to be made using ML models, while the $\Delta$-ML method might be favored, the MF$\Delta$ML method is shown to be more efficient.

\end{abstract}

\section{Introduction}

Simultaneous progress in both quantum chemistry (QC) theory and machine learning(ML) methods has resulted in a wide range of applications ranging from molecular dynamics to alloy design \cite{Haese2016MLexcitonic, Haese2016MLexcitonic, khatamsaz_multi-objective_2023, Sergei21_Chem_review_NNML}. Such ML models are often optimized to predict singular molecular properties such as atomization energies \cite{blum_QM7_dataset, zasp19a} or molecular dipole moments \cite{Veit_moleculardipoles_2020}, however also covering potential energy surfaces at both ground and excited states \cite{dral2020hierarchical, Westermayr2020review, butler_davies_review_2018}. 

\begin{figure*}[htb!]
    \centering
    \includegraphics[width=\linewidth]{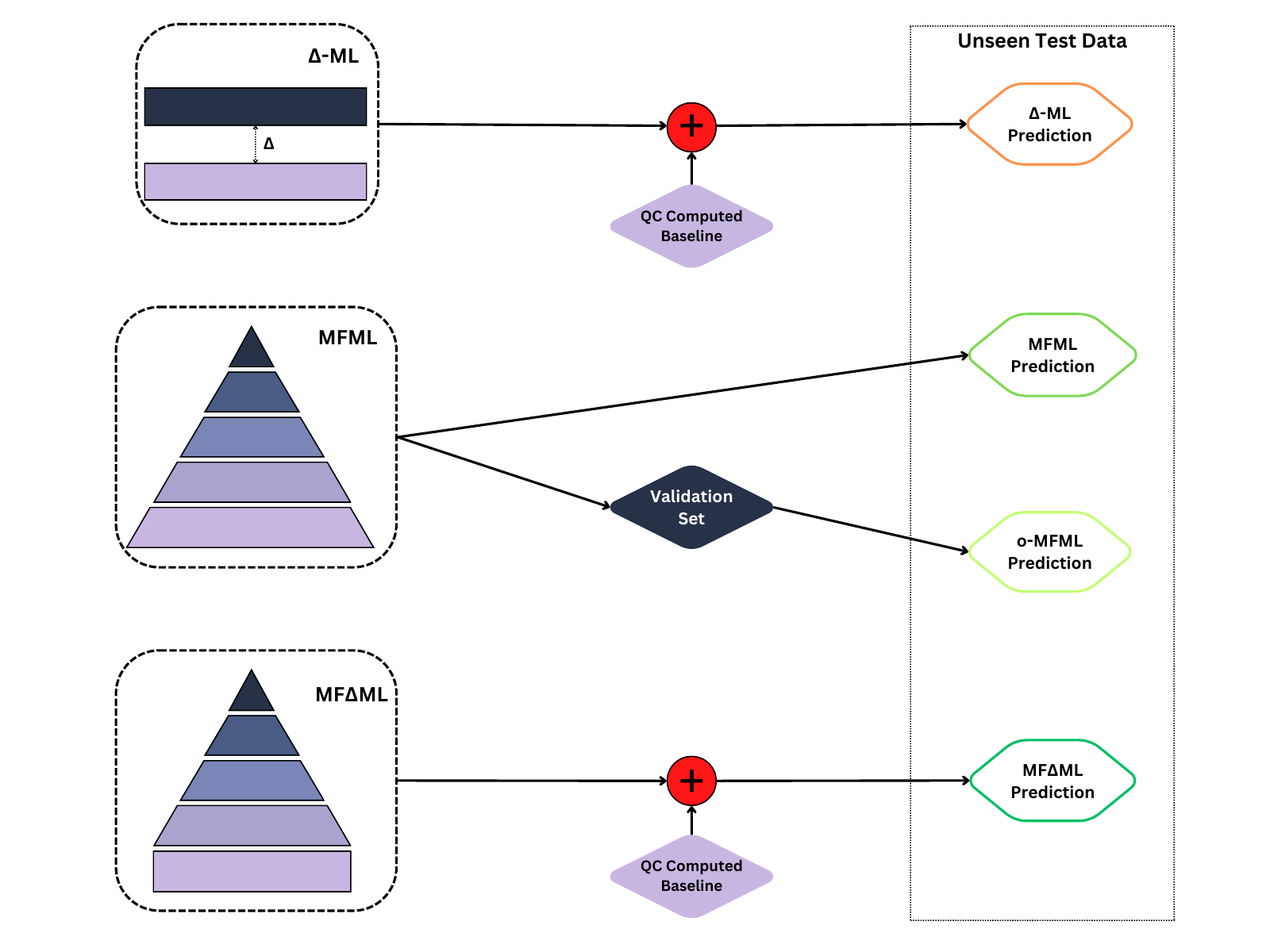}
    \caption{A visual depiction of the different ML methods benchmarked in this work. The MFML and o-MFML models do not need any further QC-calculations for the unseen test set, once they have been trained. 
    In contrast, $\Delta$-ML method and the MF$\Delta$ML method that is introduced herein, both require additional QC computations at the QC-baseline that is used in these models.
    This work benchmarks these different models to understand the time-cost versus model accuracy efficiency.
    }
    \label{fig_conceptualization}
\end{figure*}

Once trained, ML models for QC are capable of reducing the time-cost of making new predictions for unseen data tremendously, in comparison to conventional QC computation \cite{westermayr_2021_perspective, Sergei21_Chem_review_NNML, Westermayr2020review}. However, it is a common observation in ML-QC that the more samples a ML model is trained on, the better the accuracy of the model \cite{Westermayr2020review, Westermayr2020NNKRR, dral21a}.
Recent research in developing ML methods for QC has therefore begun posing a slightly different question to this entire approach: how long does it take to generate training data for such ML models? Various methodological improvements have since been developed to tackle this specific aspect of the ML-QC approach. These include $\Delta$-ML \cite{Ramakrishnan2015}, multifidelity machine learning \cite{zasp19a}, and hierarchical-ML \cite{dral2020hierarchical}. The $\Delta$-ML approach trains an ML model on the difference between two \textit{fidelities} of QC data. The term fidelity refers to the accuracy of the QC method. It is generally the case that a higher fidelity is associated with a higher compute-cost. Thus, $\Delta$-ML trains on the difference of a costly fidelity, called the \textit{target fidelity}, and a cheaper fidelity, referred to herein as the \textit{QC-baseline fidelity}. This is pictorially depicted at the top of Figure \ref{fig_conceptualization}. 
The final predictions involve making the QC-baseline fidelity calculations and adding to that the difference predicted by the $\Delta$-ML model. The h-ML approach builds several $\Delta$-ML like models for more than two fidelities where the training samples at each fidelity are decided by minimizing a cost function based on user defined error and compute-cost budget in generating training data. This model was shown to be effective 
%been shown effective 
in predicting the potential energy surface of the ground state in the $\rm CH_3Cl$ molecule \cite{dral2020hierarchical}.

Multifidelity machine learning (MFML) was introduced as a systematic generalization of the $\Delta$-ML method \cite{zasp19a} with several ML models, called \textit{sub-models}, being built with different fidelities and number of training samples. The method provides several methodological improvements over the $\Delta$-ML approach in that the cheaper fidelity is no longer calculated but are predicted within the MFML method. In both MFML and $\Delta$-ML, the training samples needed at the cheaper fidelities include the training samples used at the target fidelity. This property of the training data is referred to as the nestedness of training data \cite{zasp19a, vinod23_MFML}, that is, the training data is homogeneous.
A recent development over the MFML method is the optimized MFML (o-MFML) method \cite{vinod_2024_oMFML} which uses a validation set computed at the target fidelity to optimally combine the sub-models of MFML. This method was shown to be superior in prediction of excitation energies and atomization energies in ref.~\citenum{vinod_2024_oMFML} and shown to be better in use cases where the training data is heterogeneous, or non-nested \cite{Vinod_2024_nonnested}. The MFML and o-MFML methods are depicted in the middle of FIG.~\ref{fig_conceptualization}. Note that the conventional MFML method only needs the training data, while the o-MFML method requires the validation set computed at the target fidelity. The prediction is made without any further computation of the baseline fidelities.
Another flavor of ML methods that has recently been introduced is the use of multitask Gaussian processes (MTGPR) to predict molecular properties such as ionization potentials with a training data cost reduction of almost an order of magnitude when coupled with the $\Delta$-ML approach \cite{fisher2024multitask}.
Yet another multifidelity approach that was recently introduced is the minimal multi-level scheme which optimizes a loss function in association with a loss function to arrive at optimal cost-error balance of the multifidelity model \cite{Heinen_2024_M3L}.

This work provides a time-cost versus model accuracy benchmark of the $\Delta$-ML, MFML, and o-MFML methods. 
A uniform assessment requires that 
these models are evaluated on the same dataset. For this purpose, this work uses the QeMFi dataset \cite{vinod_2024_QeMFi_zenodo_datatset, vinod2024QeMFi_paper}, which contains five fidelities of QC properties for nine diverse molecules. In this work, the ground state energies, the first and second vertical excitation energies, and the magnitude of the electronic contribution to molecular dipole moments are used to stuyd time-cost efficiency in the multifidelity models.
In addition, the compute cost of each fidelity for each molecule is given. This enables a uniform assessment of the data efficiency, that is, the cost of generating training data for a certain model accuracy. 
In addition to the benchmarks for the $\Delta$-ML, MFML, and o-MFML models, this work introduces another multifidelity method in interest of reducing the training data cost for the ML-QC pipeline. 
This is the multifidelity $\Delta$-ML (MF$\Delta$ML) method wherein, a MFML model is built with several $\Delta$-ML models which predict the QC property for different fidelities. 
This method is depicted at the bottom of FIG.~\ref{fig_conceptualization}. The QC-baseline fidelity is used to create the $\Delta$-ML models of the higher fidelities in the training data structure. The final prediction requires the calculation of the QC-baseline fidelity to which the prediction from the MF$\Delta$ML model are appended. Further details of this method are presented in section~\ref{MFdelML_method}. This method is also benchmarked alongside the $\Delta$-ML, MFML, and o-MFML methods. 

The structure of the rest of the manuscript is as follows. The QeMFi dataset and the different ML methods that are benchmarked on the dataset are detailed in section \ref{Methods}. Details of the newly introduced MF$\Delta$ML method are also explained therein. 
The results for the benchmarks are then explained in section \ref{results} for the diverse models that are studied including MFML and MF$\Delta$ML. 
The results are discussed in detail with inferences that can be drawn from them. Finally, a discussion on the work presented here is accompanied by future outlook for the implementation of multifidelity methods for QC. 

\section{Methods}\label{Methods}
This section explains the methodological pre-requisites to understanding the results of section \ref{results}. The dataset used in the main manuscript is detailed along with the different ML methods that are benchmarked for efficiency in this work.
\subsection{QeMFi Dataset}\label{dataset_method}
The various ML models built and assessed in this work are benchmarked on the QeMFi dataset \cite{vinod_2024_QeMFi_zenodo_datatset, vinod2024QeMFi_paper}. This dataset contains 135,000 geometries of nine chemically diverse molecules, each with several chemical conformers. For each geometry, five fidelities are calculated with the TD-DFT formalism with increasing basis set sizes: STO-3G, 3-21G, 6-31G, def2-SVP, and def2-TZVP. For the rest of this work, these fidelities are referred to by their shorthand such as SVP or TZVP. 

\begin{figure*}[htb!]
    \centering
    \includegraphics[width=0.75\linewidth]{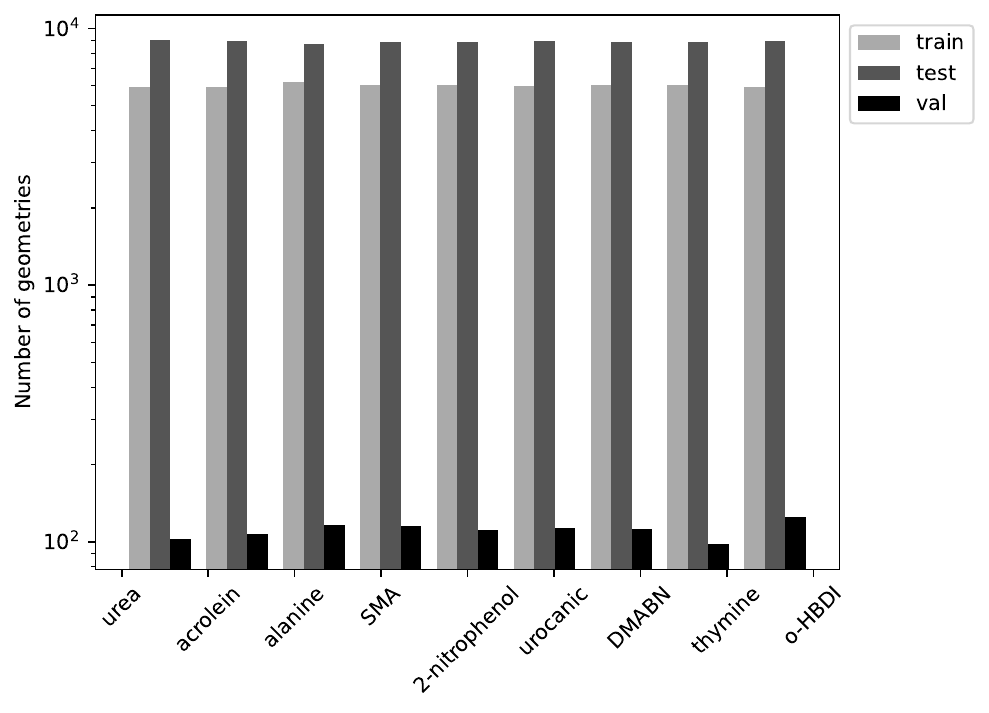}
    \caption{Distribution of training, validation, and test sets used in this work. All the nine molecules of the QeMFi dataset are evenly present in each of the sets. The same train/test/validation split is used for all QC properties studied in this work.}
    \label{fig_sampling_distr}
\end{figure*}
In this work, the different ML models are evaluated for the following QC properties of the QeMFi dataset:
\begin{itemize}
    \item ground state energies, 
    \item first vertical excitation energies ($E_{(1)}$), 
    \item second vertical excitation energies ($E_{(2)}$), 
    \item magnitude of electronic dipole moments ($\lvert\boldsymbol{\mu}_e\rvert$)
\end{itemize} 
Although ground state energies are considered easier than excitation energies for ML models to predict \cite{Westermayr2020review}, both of these are studied in this work to ensure that the efficiency analysis of the multifidelity methods is less dependent of the QC property. 
The magnitude of the electronic contribution to molecular dipole moments is studied instead of a component-wise vector prediction of the property since the latter requires specialized molecular representations which are equivariant under rotation and translation \cite{Veit_moleculardipoles_2020} unlike for the case of energies where invariance is mandated \cite{lilienfeld_2015_descriptors_requirement}. This is since the vector quantity of dipole moments is dependent on the orientation of the molecule itself. On the other hand, the magnitude is a rotation and translation invariant QC property which can be modeled with most conventional molecular descriptors such as the Coulomb Matrices used in this work (see section \ref{CM_method}). The use and development of specialized descriptors for dipole moments lies outside the aim of this work, which is focused on the efficiency of the multifidelity methods. Furthermore, only the electronic contribution is studied since the nuclear contribution to the molecular dipole moment is identical for a molecule regardless of the fidelity of data used. 

The split of data into training, validation, and test sets was uniform across these different QC properties. Of the 135,000 geometries, a random collection of 54,000 data points were removed to be used for the necessary training data sets. From the remaining geometries, 1,000 samples were chosen at random to be the validation set to be used for the optimization procedure in o-MFML. The remaining 85,000 samples were set aside as the test set. It is to be noted that the test set is not used at any stage of training the models and is therefore a true proxy for unseen data. 
Since the QeMFi dataset consists of nine molecules of different sizes and chemical complexities, one should make sure that the different molecules are sufficiently represented in the training and test sets. In interest of this sanity check,
the composition of the training, validation, and test sets is shown in FIG.~\ref{fig_sampling_distr} based on this selection choice. 
It can be seen that the nine molecules are uniformly represented in all three sets. This form of sampling ensures that there is no separate influence of the composition of the dataset itself.

\subsection{Molecular Descriptor} \label{CM_method}
Unsorted Coulomb Matrices (CM) \cite{Rup12CM, david2020moleculardescriptors} are used in this work as molecular descriptors or \textit{representations}.
For a molecule, the entries of CM, $C$, are computed as
\begin{equation}
C_{i,j}:=
    \begin{cases}
        \frac{Z_i^{2.4}}{2}~,&i=j\\
        \frac{Z_i\cdot Z_j}{\left\lVert \boldsymbol{R}_i-\boldsymbol{R}_j\right\rVert}~,&i\neq j~,
    \end{cases}
    \label{CM_eq}
\end{equation}
where the Cartesian coordinate of the $i$-th atom is $\boldsymbol{R}_i$ with $Z_i$ being the atomic charge. The indices $i,j$ run over the atoms of the molecule. Note that the CM are symmetric matrix representations. That is, only the upper triangular entries of the matrix $C$ would be unique. 
Thus, for a molecule consisting of $m$ atoms, the number of unique entries of the corresponding CM are $m(m+1)/2$. 
Since the molecules have different sizes, the molecular descriptors are padded by zeros to maintain a uniform size of the CM. The largest molecule in the QeMFi dataset is o-HBDI with 22 atoms resulting in the padded CM size of 253 entries. 
Some literature uses a row-norm sorted CM as molecular representation, however it is avoided here since this form of sorting is known to produce discontinuities in the representation  \cite{Rup12CM, krae20a}. 

\subsection{Kernel Ridge Regression} \label{KRR_method}
Given a molecular descriptor $\boldsymbol{X}_q$, the KRR model $P^{(f)}_{\rm KRR}$, for a fidelity $f$, predicts a QC property, say ground state energy, as
\begin{equation}
    P^{(f)}_{\rm KRR}\left(\boldsymbol{X}_q\right) := \sum_{i=1}^{N^{(f)}_{\rm train}} \alpha^{(f)}_i k\left(\boldsymbol{X}_q,\boldsymbol{X}_i\right)~,
    \label{eq_KRR_def}
\end{equation}
where $k(\cdot,\cdot)$ is the kernel function that adds non-linearity to the model, and $N_{\rm train}^{(f)}$ is the number of training samples used at some fidelity $f$.
This work uses the Mat\'ern kernel given by
\begin{equation}
    k\left(\boldsymbol{X}_i,\boldsymbol{X}_j\right) = \exp{\left(-\frac{\sqrt{3}}{\sigma}\left\lVert \boldsymbol{X}_i-\boldsymbol{X}_j\right\rVert_2^2\right)}  \times \left(1+\frac{\sqrt{3}}{\sigma}\left\lVert \boldsymbol{X}_i-\boldsymbol{X}_j\right\rVert_2^2\right)~,
    \label{eq_matern}
\end{equation}
where $\sigma$ is a length scale hyperparameter. This work manually sets the value of $\sigma$ to 150.0 for ground state energies, 20.0 for excitation energies, and 3000.0 for the magnitude of electronic contribution to molecular dipole moments based on a hyper-parameter grid search. 
The vector $\boldsymbol{\alpha}^{(f)}$ of coefficients of the KRR model is calculated by solving a system of equations $(\boldsymbol{K}+\lambda \boldsymbol{I}) \boldsymbol{\alpha}^{(f)} = \boldsymbol{y}^{(f)}$. Here, $\boldsymbol{K} = \left(k(\boldsymbol{X}_i,\boldsymbol{X}_j)\right)_{i,j=1}^{{N}_{\rm train}}$ is called the kernel matrix, $\boldsymbol{I}$ the identity matrix, and $\boldsymbol{y}^{(f)} = \left(y^{(f)}_1, y^{(f)}_2, \ldots, y^{(f)}_{{N}_{\rm train}}\right)^T$ is the vector of QC properties from the training set, which can be denoted as $\mathcal{T}^{(f)}$. 
The regularization parameter $\lambda$ penalizes overfitting and is set in this work to be $10^{-10}$ for all applications. 

\subsection{\texorpdfstring{$\Delta$}{Delta}-ML Approach} \label{delML_method}
The $\Delta$-ML approach uses training data computed at two different fidelities \cite{Ramakrishnan2015} to train a model. Consider a training set computed at some fidelity $F$ given as $\mathcal{T}^F:=\{(\boldsymbol{X}_i, y_i^F)\}_{i=1}^{N_{\rm train}^F}$ 
where $\boldsymbol{X}_i$ are the molecular descriptors and the corresponding QC property are $y_i^F$. For the same molecular descriptors, consider a set of QC computational calculations made at some lower fidelity $QC_b<F$ given as $\mathcal{T}^{QC_b}:=\{(\boldsymbol{X}_i,y_i^{QC_b})\}_{i=1}^{N_{\rm train}^{F}}$. Then the $\Delta$-ML approach trains a ML model to learn the difference of the property between the two fidelities, that is $\Delta_{\rm QC_b}^F=\boldsymbol{y}^F-\boldsymbol{y}^{QC_b}$. Hereon, $F$ is referred to as the target fidelity. 
The final prediction of a KRR model with the $\Delta$-ML approach is to first predict the difference between the target and baseline fidelities and then to add the QC calculation of the baseline fidelity for the evaluation samples. In other words, for a given query representation, $\boldsymbol{X}_q$, the $\Delta$-ML prediction is given as
\begin{equation}
    P_{\Delta}^{F;QC_b}:= P_{\rm KRR}^{\Delta_{\mathrm{QC_b}}^F}(\boldsymbol{X}_q) + y^{QC_b}_q~,
    \label{delML_pred}
\end{equation}
where, $P_{\rm KRR}^{\Delta_{\mathrm{QC_b}}^F}$ is the KRR prediction of the difference, and $y^{\mathrm{QC_b}}_q$ is the QC computation of the baseline fidelity for the query molecule.
Throughout this work, since the target fidelity is always def2-TZVP, the various $\Delta$-ML models are referred to by their QC-baseline fidelities. 

\subsection{Multifidelity Machine Learning Method} \label{MFML_methods}
Multifidelity ML (MFML) is a systematic generalization of the $\Delta$-ML method, which iteratively combines KRR \textit{sub-models} built at more than two fidelities between the target  fidelity $F$ and some baseline fidelity $f_b$ \cite{zasp19a, vinod_2024_oMFML}. The sub-models are identified by a composite index that denotes the fidelity, $f$, and the number of training samples used in that fidelity, $2^{\eta_f}=N_{\rm train}^{(f)}$. This composite index is then represented as $\boldsymbol{s}=(f,\eta_f)$. The MFML model can then be written as 
\begin{equation}
    P_{\rm MFML}^{(F,\eta_F;f_b)}\left(\boldsymbol{X}_q\right) := \sum_{\boldsymbol{s}\in\mathcal{S}^{(F,\eta_F;f_b)}} \beta_{\boldsymbol{s}} P^{(\boldsymbol{s})}_{ \rm KRR}\left(\boldsymbol{X}_q\right)~.
    \label{eq_MFML_linearsum}
\end{equation}
The summation runs over the set of chosen sub-models for MFML 
$$\mathcal{S}^{(F,\eta_F;f_b)}:= \Big{\{}(f,\eta_f) \lvert
    f \in\left\{f_b,\ldots,F\right\},
     \eta_f\in\{\eta_F,\ldots,2^{F-f_b}\cdot \eta_F \},
    F+\eta_F-1\leq f+\eta_f \leq F+\eta_F \Big{\}}~.
$$ 
The sub-models are chosen based on $f_b$ and $N_{\rm train}^{(F)}=2^{\eta_F}$, that is the number of training samples used at the highest fidelity.

The $\beta_{\boldsymbol{s}}$ from Eq.~\eqref{eq_MFML_linearsum} are the coefficients of the linear combination of these sub-models. 
Conventional MFML uses the following value of $\beta_{\boldsymbol{s}}$ as prescribed in ref.~\citenum{zasp19a}:
\begin{equation}
    \beta_{\boldsymbol{s}}^{\rm MFML} = \begin{cases}
        +1, & \text{if } f+\eta_f = F+\eta_F\\
        -1, & \text{otherwise}
    \end{cases}~.
    \label{eq_MFML_beta_i}
\end{equation}
As an example, consider the case for a three fidelity model, that is $f_b=1$, $F=3$, with $2^2$ training samples used at $F$. Here $\eta_F = 2$, thereby giving:
$$
P_{\rm MFML}^{(3,2;1)}:= P_{\rm KRR}^{(3,2)} - P_{\rm KRR}^{(2,2)} + P_{\rm KRR}^{(2,3)} - P_{\rm KRR}^{(1,3)} + P_{\rm KRR}^{(1,4)}~.
$$
The individual KRR models identified by the composite index, $\boldsymbol{s}$, are chosen per $\mathcal{S}^{(F,\eta_F;f_b)}\equiv\mathcal{S}^{(3,2;1)}$. The $\pm1$ as $\beta_{\boldsymbol{s}}$ is set based on Eq.~\eqref{eq_MFML_beta_i}. 

Ref.~\citenum{vinod_2024_oMFML} introduced \textit{optimized MFML} (o-MFML) for the prediction of QC properties. This method allows for values of $\boldsymbol{\beta}_s$ different from those in Eq.~\eqref{eq_MFML_beta_i} through an optimization process over a validation set, say,
$\mathcal{V}^F_{ \rm val}:=\{(\boldsymbol{X}_q^{ \rm val},y^{\rm val}_q)\}_{q=1}^{N_{ \rm val}}$. 
Given a target fidelity $F$, with $N^{(F)}_{\rm train}=2^{\eta_F}$ training samples and a given baseline fidelity $f_b$, the prediction of the o-MFML model can be defined as 
\begin{equation}
    P_{\rm o-MFML}^{\left(F,\eta_F;f_b\right)}\left(\boldsymbol{X}_q\right) := 
    \sum_{\boldsymbol{s}\in \mathcal{S}^{(F,\eta_F;f_b)}}\beta_{\boldsymbol{s}}^{\rm opt} P^{(\boldsymbol{s})}_{\rm KRR} \left(\boldsymbol{X}_q\right)~,
    \label{eq_POM_def}
\end{equation}
where $\beta_{\boldsymbol{s}}^{\rm  opt}$ are the optimized coefficients that are obtained by solving the optimization problem
\begin{equation*}
    \beta_{\boldsymbol{s}}^{\rm opt} =\arg\min_{\beta_{\boldsymbol{s}}} 
    \left\lVert \sum_{v=1}^{N_{ val}} \left(y_v^{\rm val} - \sum_{\boldsymbol{s}\in S^{(F,\eta_F;f_b)}} \beta_{\boldsymbol{s}} P^{(\boldsymbol{s})}_{\rm KRR}\left(\boldsymbol{X}^{ val}_v\right)\right) \right\rVert_p~,
\end{equation*}
where one minimizes some $p$-norm on the validation set defined beforehand. The optimization procedure is carried out using ordinary least squares (OLS) as discussed in ref.~\citenum{vinod_2024_oMFML}. OLS explicitly uses a $p=2$ norm.

\subsection{Multifidelity \texorpdfstring{$\Delta$}{Delta}-Machine Learning Method} \label{MFdelML_method}
Given the development of MFML for an ordered hierarchy of fidelities, $f\in\{1,2,\ldots,F\}$, one can consider a case where all the training energies are `centered' by the energies of the lowest fidelity, $f=1$. This approach essentially creates a $\Delta$-ML model over the MFML model and can be termed as multifidelity $\Delta$-machine learning (MF$\Delta$ML) method. The prediction using this method for a query representation $\boldsymbol{X}_q$ is given as:
\begin{equation}
    P_{\mathrm{MF}\Delta\mathrm{ML}}^{F,\eta_F;f_b,QC_b}(\boldsymbol{X}_q):= \sum_{\boldsymbol{s}\in\mathcal{S}^{(F,\eta_F;f_b)}} \beta_{\boldsymbol{s}} P^{(\boldsymbol{s})}_{\Delta}\left(\boldsymbol{X}_q\right)~.
    \label{eq_MFdelML}
\end{equation}
Here, $P_{\Delta}^{(\boldsymbol{s})}$ are $\Delta$-ML models identified by Eq.~\eqref{delML_pred} where $QC_b$ would be the fidelity $f=1$, the target fidelity for these $\Delta$-ML models would be fidelity $f$. The MF$\Delta$ML model is built for some baseline fidelity $f_b>1$ for a target fidelity $F$. 
Thus, with the MF$\Delta$ML approach, the multifidelity model predicts the difference in energies of some fidelity $f$ and the QC-baseline, $f_b^{\rm QC}$. 
With this definition one can also readily extend the concept of o-MFML to optimized MF$\Delta$ML (o-MF$\Delta$ML). This work benchmarks the time-cost efficiency of both MF$\Delta$ML and o-MF$\Delta$ML contrasted with the single fidelity, $\Delta$-ML, and MFML methods described above. 

\subsection{Error Metric}
Model error for the various ML models created in this work are reported as mean absolute error (MAE). MAE are calculated using a discrete $L_1$ norm as:
\begin{equation}
    \mathrm{MAE} = \frac{1}{N_{\rm test}}\sum_{q=1}^{N_{\rm test}}\left\lvert P_{ \rm ML}\left(\boldsymbol{X}_q^{ \rm test}\right) - {y}^{\rm test}_q\right\rvert~.
    \label{eq_MAE}
\end{equation} 
The model $P_{\rm ML}$ in Eq.~\eqref{eq_MAE}, can be either the single fidelity KRR or the different MFML and $\Delta$-ML models used in this work. The MAE are reported using learning curves, which are a measure of model error versus model complexity \cite{cortes1993learning, muller1996numerical_LearningCurve}. Although, it is to be noted that in some cases such as training neural network models, learning curves depict model error versus training epochs. In general one anticipates that the model error decreases with increasing model complexity. Since this work uses KRR, the model complexity is directly described by the number of training samples used for the KRR model. Therefore, this work reports the learning curves as MAE versus the number of training samples used to build the ML model. For multifidelity methods and $\Delta$-ML methods, the training samples used at the target fidelity are reported. 
The model error for ground state energies and excitation energies are reported in units of kcal/mol while the magnitude of dipole moments is reported in units of Debye. 
In order to benchmark the models studied in this work, model error as a function of time-cost to generate the training data are analyzed. The time-cost is calculated for all the training data that is needed for the specific model. 
For MFML, the time-cost considers the cost of generating the training data for all the fidelities. For o-MFML, in addition to the cost of the multifidelity training data, the time-cost includes the cost of the validation set computed at the target fidelity of TZVP.
For $\Delta$-ML, the time cost is the cost of the training data at the target fidelity and baseline fidelity in addition to the cost of making the QC-baseline calculations for the test set.
For the case of MF$\Delta$ML, the cost of the multifidelity training data along with the cost of the QC calculations for the test set is considered.

\section{Results}\label{results}
To assess the various models described in section \ref{Methods}, the models were trained on a pool of multifidelity data taken from the QeMFi dataset \cite{vinod_2024_QeMFi_zenodo_datatset, vinod2024QeMFi_paper}.They are evaluated on a holdout test set sampled as explained in section~\ref{dataset_method}. First, learning curves are studied to understand models accuracy as a function of the number of training samples used at the costliest fidelity, that is TZVP. 
Next, the cost of training the multifidelity models is calculated including any additional costs such as the validation set cost for o-MFML, or QC-baseline computation for the $\Delta$-ML variants. This is the benchmark that is presented in this work. The time-cost of using the different multifidelity methods are studied with recommendations of which method to use when based on the results. 

\subsection{Learning Curves}
\begin{figure*}[htb!]
    \centering
    \includegraphics[width=0.8\textwidth]{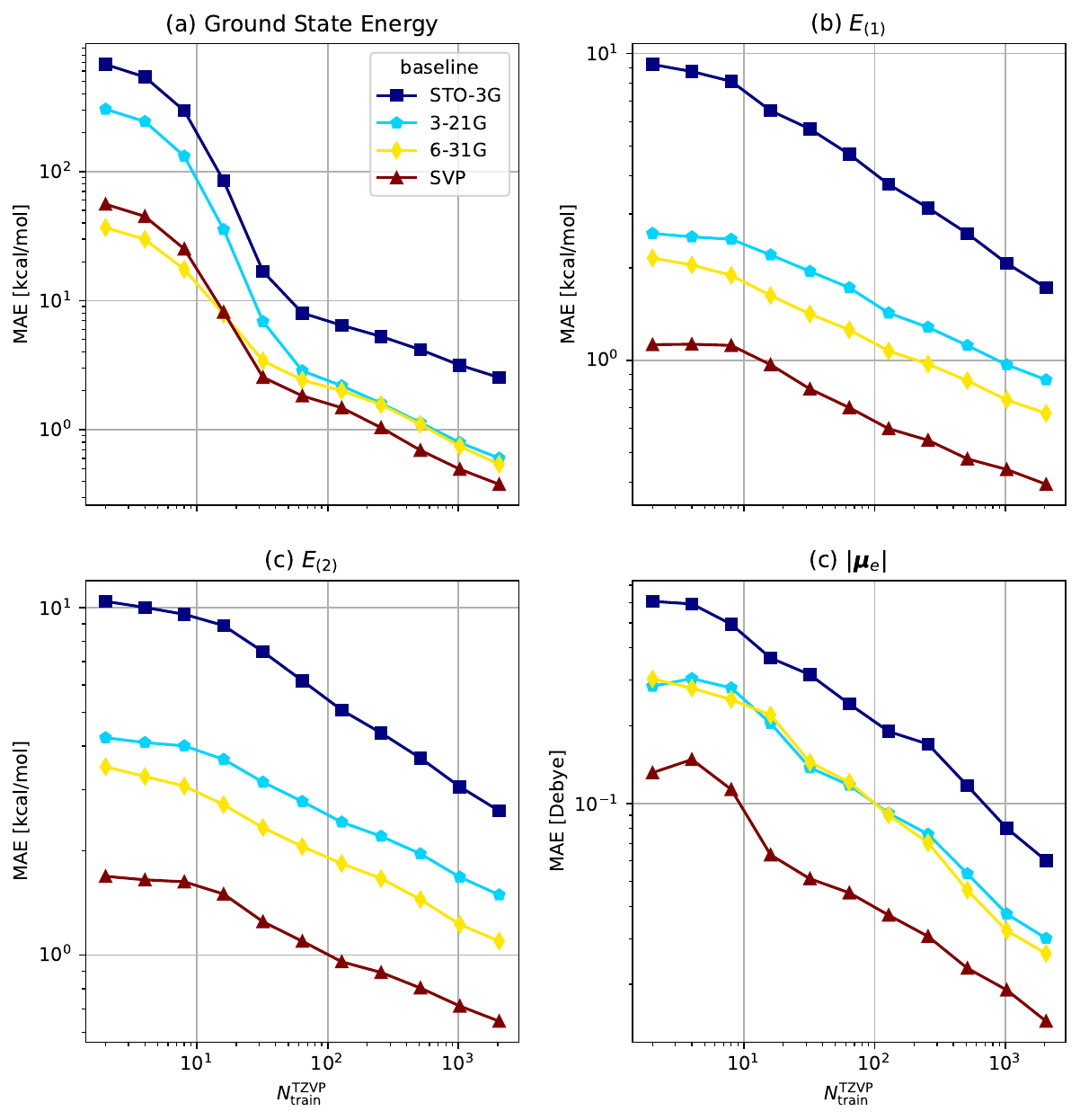}
    \caption{Learning curves for $\Delta$-ML with varying $QC_b$. These are shown for the prediction of ground state energies, first vertical excitation energies ($E_{(1)}$), second vertical excitation energies ($E_{(2)}$), and the magnitude of electronic contribution to molecular dipole moments ($\lvert\boldsymbol{\mu}_e\rvert$). Across the QC properties, it is observed that the closer the $QC_b$ is in hierarchy to the target fidelity, the better the model accuracy, as also observed in ref.~\citenum{Ramakrishnan2015}.}
    \label{fig_DeltaML_LC}
\end{figure*}

\begin{figure*}[htb!]
    \centering
    \includegraphics[width=0.8\textwidth]{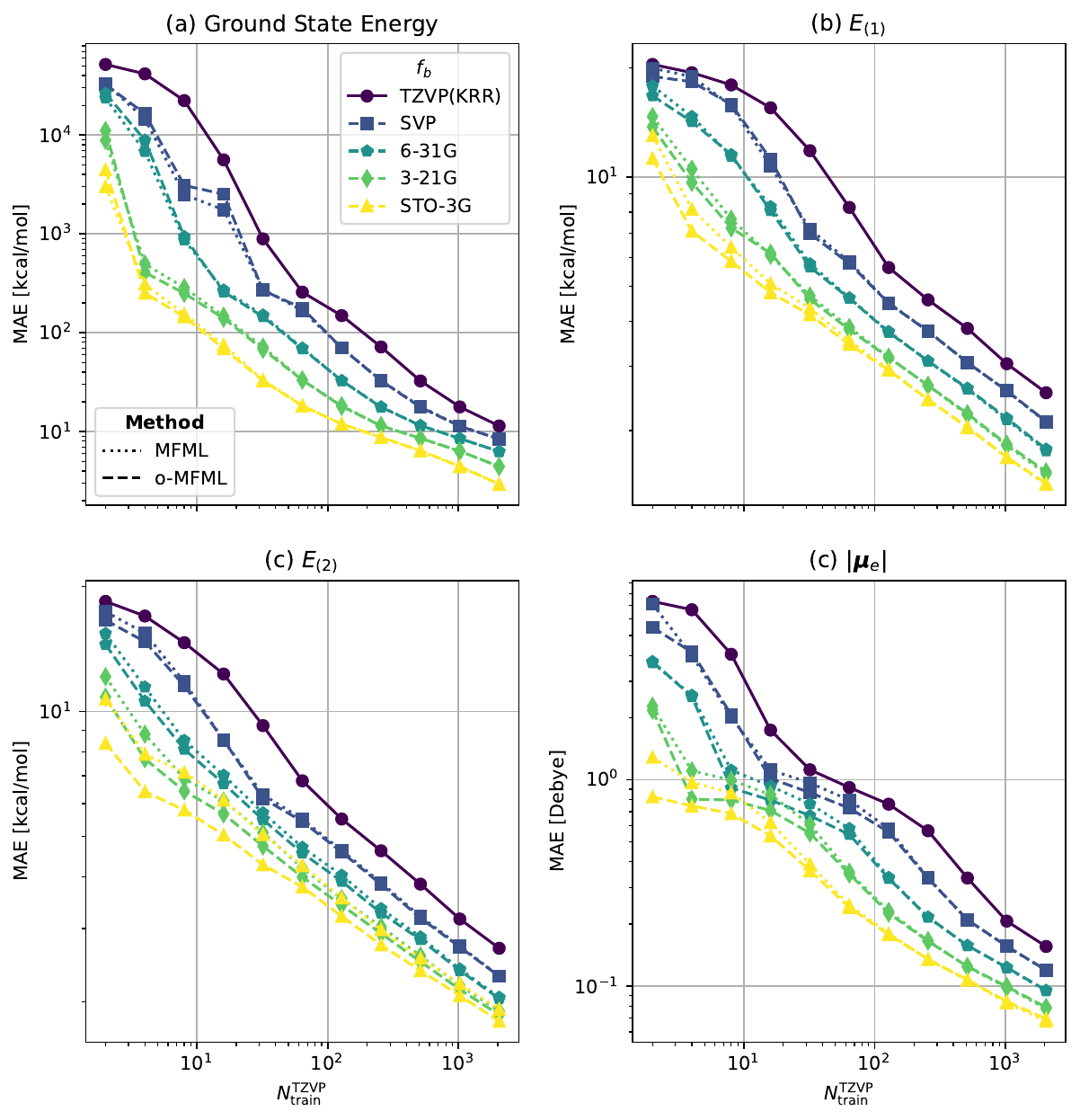}
    \caption{MFML and o-MFML learning curves for different QC properties studied in this work. The different baselines used in the MFML and o-MFML models are reported in the legend. It is seen that the o-MFML model does not provide a significant improvement over the conventional MFML model in terms of MAE. This could indicate that the MFML combination of sub-models is already sufficiently optimized.}
    \label{fig_MFML_LC}
\end{figure*}

\begin{figure*}[htb!]
    \centering
    \includegraphics[width=0.8\textwidth]{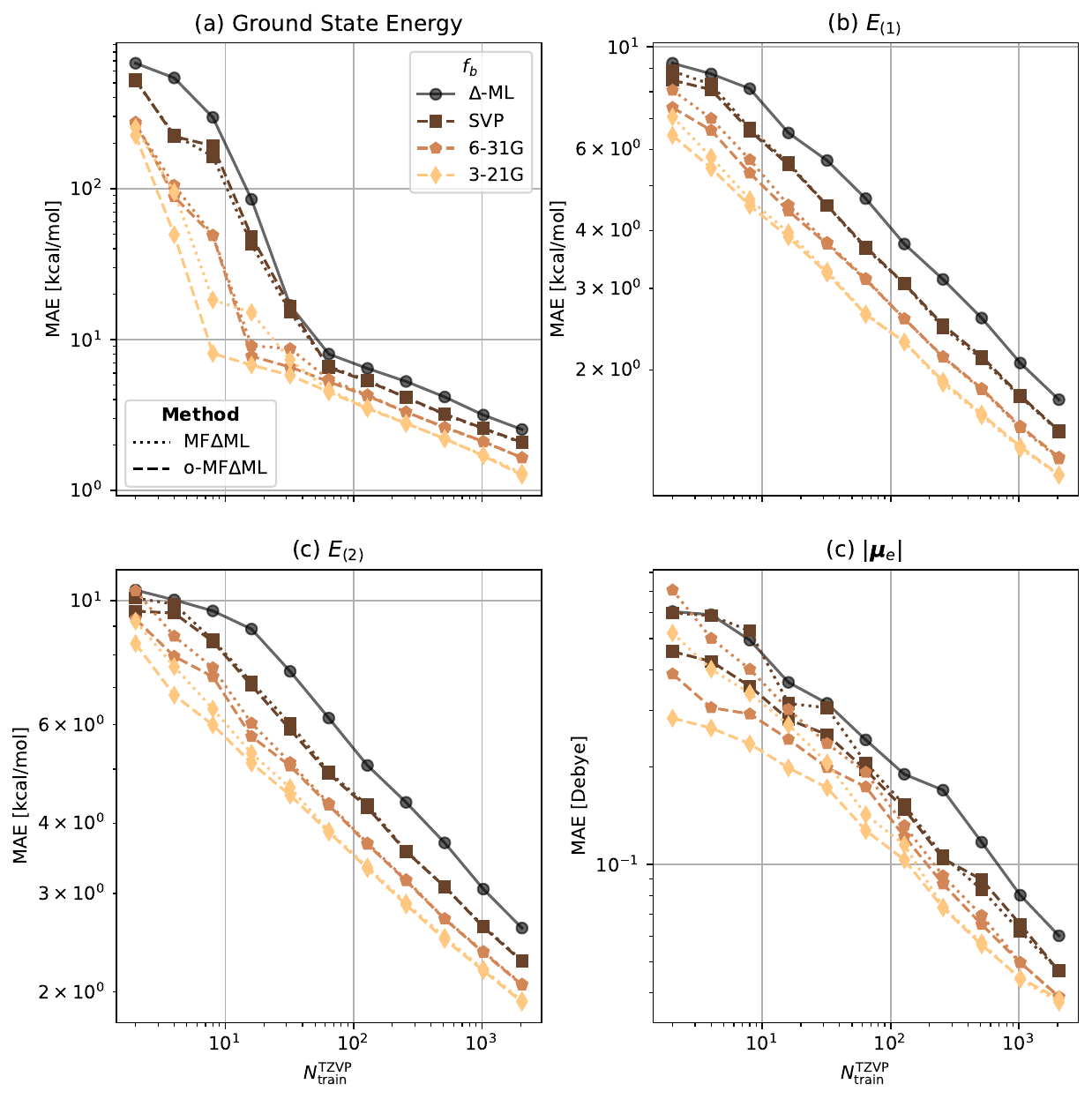}
    \caption{Learning curves for the newly introduced MF$\Delta$ML and o-MF$\Delta$ML with differing baseline fidelities. }
    \label{fig_deltaMFML_LC}
\end{figure*}

FIG.~\ref{fig_DeltaML_LC} reports the learning curves for the $\Delta$-ML method with varying baseline fidelities for the (a) prediction of ground state energies, (b) first excitation energies ($E_{(1)}$), (c) second excitation energies ($E_{(1)}$), and (d) the magnitude of electronic contribution to the molecular dipole moment ($\boldsymbol{\mu}_e$). For all QC properties, one notices that the use of cheaper $QC_b$ offsets the learning curves upwards. In other words, the closer $QC_b$ is to the target fidelity of $F$, the better the $\Delta$-ML model is. 
The difference in MAE between the STO-3G baseline and the SVP baseline for $\Delta$-ML is observed to be almost an order of magnitude. This observation is similar to what is made in ref.~\citenum{Ramakrishnan2015}, where the $\Delta$-ML approach was first introduced. 
In the supporting information (SI), the results of similar experiment for the QM7b dataset \cite{montavon2013machine} are reported. In that experiment, different QC levels of theory are considered as fidelities, as opposed to the different basis sets that are considered as fidelities, here. While the experiment is detailed in sections S1 and S2, the results once again confirm the trend of the $\Delta$-ML model resulting in lower error for a QC-baseline that is closer to the target fidelity as seen in FIG.~S1.  

FIG.~\ref{fig_MFML_LC} reports the learning curves for MFML and o-MFML models as introduced in ref.~\citenum{vinod_2024_oMFML} and explained in section \ref{MFML_methods}. Note that the vertical axis are scaled differently for each QC property with the units of the energies being reported in kcal/mol while using Debye for dipole moments. For each cheaper $f_b$, one observes that the learning curve is lowered by a constant offset. For ground state energies, as seen in FIG.~\ref{fig_MFML_LC}(a), there is a region of pre-asymptotics that is observed for small training set sizes up to $N_{\rm train}^{\rm TZVP}=16$. However, for larger training set sizes, the learning curves show constant lowered offsets and nearly constant slope on the log-scaled axes. 
For both $E_{(1)}$ and $E_{(2)}$, this behavior of the learning curves is observed even for small training set sizes. 
As seen in FIG.~\ref{fig_MFML_LC}(d), the learning curves for the prediction of $|\boldsymbol{\mu}_e|$ too show lowered offsets with the addition of cheaper baseline fidelities. In all cases, the learning curves continue to have constant negative slope for large training set sizes.
This indicates that further addition of training samples could indeed reduce model error. 
The learning curves for MFML and o-MFML do not show much difference in the model accuracy for any of the QC properties. This could be due to the small size of the validation set in comparison to the large training and test set sizes that are considered. Ref.~\citenum{vinod_2024_oMFML} proposes that such a behavior could also be due to the conventional MFML model being already optimized for the combination of the multifidelity sub-models. In such outcomes, it is to be noted that the MFML model is computationally more efficient over the o-MFML model, since there is no added cost of generating a validation set. More about this is discussed in section \ref{timecostresults} in relation with the time-cost results. 

Finally, the learning curves for the MF$\Delta$ML and o-MF$\Delta$ML methods as introduced in section \ref{MFML_methods} are presented in FIG.~\ref{fig_deltaMFML_LC} for the different QC properties studied in this work. 
In this case, the energies are offset by the STO-3G fidelity. That is, the STO-3G energies are subtracted from the energies at the other fidelities as explained in section \ref{delML_method}.
Thus the lowest baseline that is used for the MF$\Delta$ML model is the 3-21G fidelity. The learning curves are contrasted with the $\Delta$-ML model built for $F=$TZVP and $f_b$=STO-3G. All learning curves show a region of pre-asymptotics for training set sizes up to $N_{\rm train}^{\rm TZVP}=64$. In this region of pre-asymptotics, the o-MF$\Delta$ML method provides some improvement for the 3-21G baseline fidelity. Beyond that, the MF$\Delta$ML and o-MF$\Delta$ML methods result in very similar MAE values. With MF$\Delta$ML, it can be seen that the addition of cheaper fidelities to the model results in a constant lowered offset with a negative slope. The learning curves for the cheaper baselines of MF$\Delta$ML are seen to be below the conventional $\Delta$-ML model. 

\begin{table}[htb!]
    \centering
    \begin{tabular}{|l|c|c|c|c|}
    \hline 
         \textbf{Property/Model}& \textbf{KRR} & \textbf{MFML} & $\boldsymbol{\Delta}$\textbf{-ML} & \textbf{MF}$\boldsymbol{\Delta}$\textbf{ML}\\
         \hline 
         \hline 
         Ground state energies [kcal/mol] & 11.41 & 2.93 & 2.54 & 1.3\\
         $E_{(1)}$ [kcal/mol] & 2.54 & 1.42 & 1.72 & 1.18\\
         $E_{(2)}$ [kcal/mol] & 2.69 & 1.92 & 2.6 & 1.93 \\
         $\lvert\boldsymbol{\mu}_e\rvert$ [Debye] & 0.16 & 0.07 & 0.06 & 0.04 \\
         \hline
    \end{tabular}
    \caption{MAE in appropriate units for single fidelity KRR and multifidelity models with $N_{\rm train}^{TZVP}=2^{11}$ for different QC properties for STO-3G as the cheapest fidelity included. MAEs for o-MFML and o-MF$\Delta$ML are not shown since they are very close in value to the conventional MFML and MF$\Delta$ML values.}
    \label{tab_MAEs_comparison}
\end{table}

TABLE \ref{tab_MAEs_comparison} reports the MAEs of single fidelity KRR and the different multifidelity models for comparison of the model accuracies for different QC properties. 
The MAEs correspond to models built with $N_{\rm train}^{\rm TZVP}=2^{11}$. The MFML model is built with $f_b$=STO-3G, the $\Delta$-ML model with $QC_b=$STO-3G, and the MF$\Delta$ML model with $QC_b=$STO-3G and $f_b=$3-21G. These correspond to the last data point in the learning curves presented in this section. One observes in TABLE \ref{tab_MAEs_comparison} that the MF$\Delta$ML model has the lowest error regardless of the QC property that is studied. However, for $\lvert\boldsymbol{\mu}_e\rvert$ the difference is not as pronounced w.r.t.~MFML and $\Delta$-ML. The largest difference in the errors is seen for ground state energies, while both the excitation energies show considerable difference in the MAEs between single fidelity KRR and other multifidelity models. Thus, if only the model error is considered, the MF$\Delta$ML method is seen to be a methodological improvement over both MFML and $\Delta$-ML approaches.

\subsection{Time-Cost Assessment}\label{timecostresults}
\begin{figure*}
    \centering
    \includegraphics[width=0.8\textwidth]{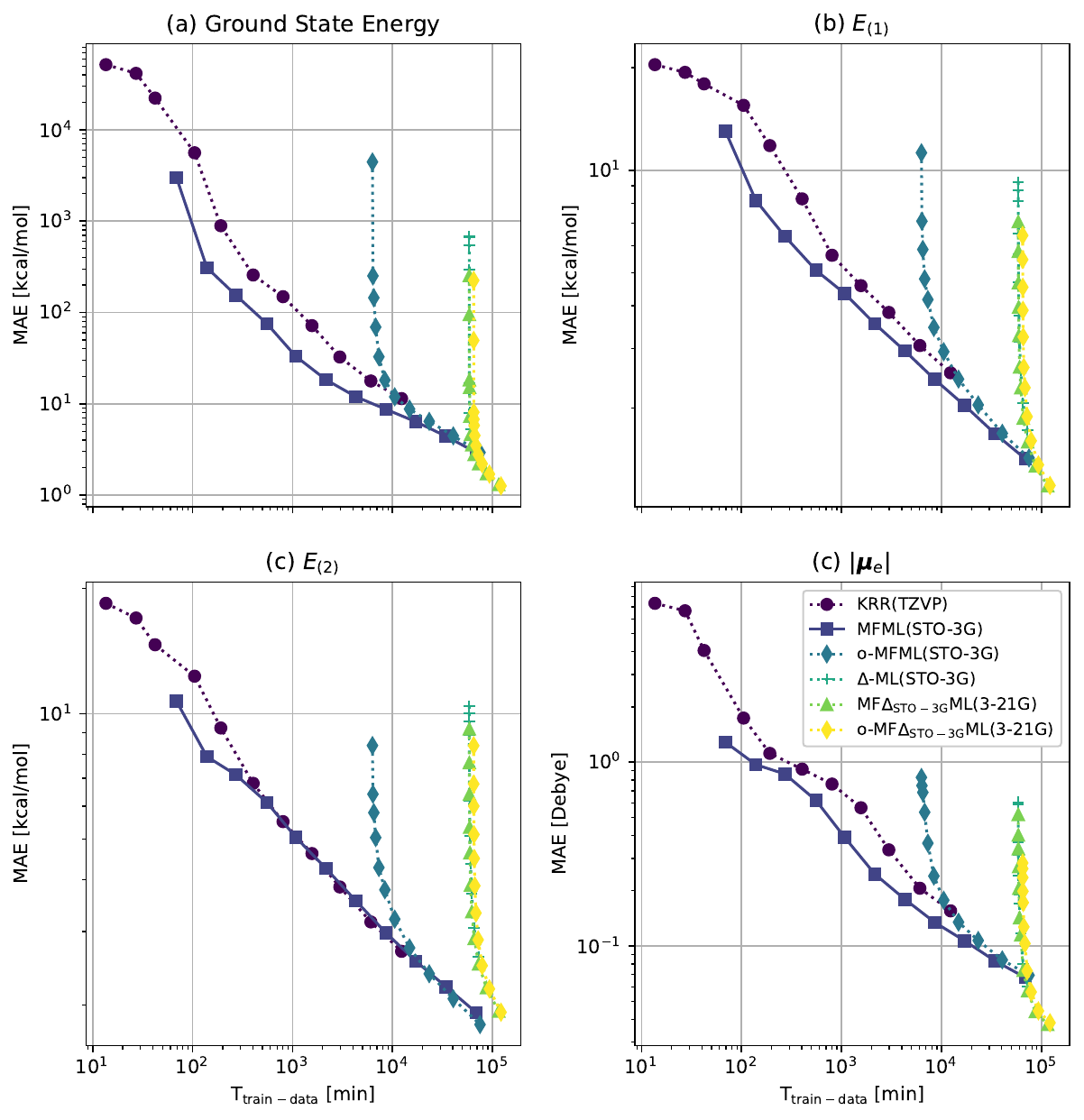}
    \caption{Time-cost assessment of the different multifidelity models for diverse QC properties. The x-axis reports $T_{\rm train-data}$ in minutes which is the time taken to generate training data. For $\Delta$-ML models this also includes the cost of the QC-baseline calculations.}
    \label{fig_time_all}
\end{figure*}

\begin{figure*}[htb!]
    \centering
    \includegraphics[width=0.8\textwidth]{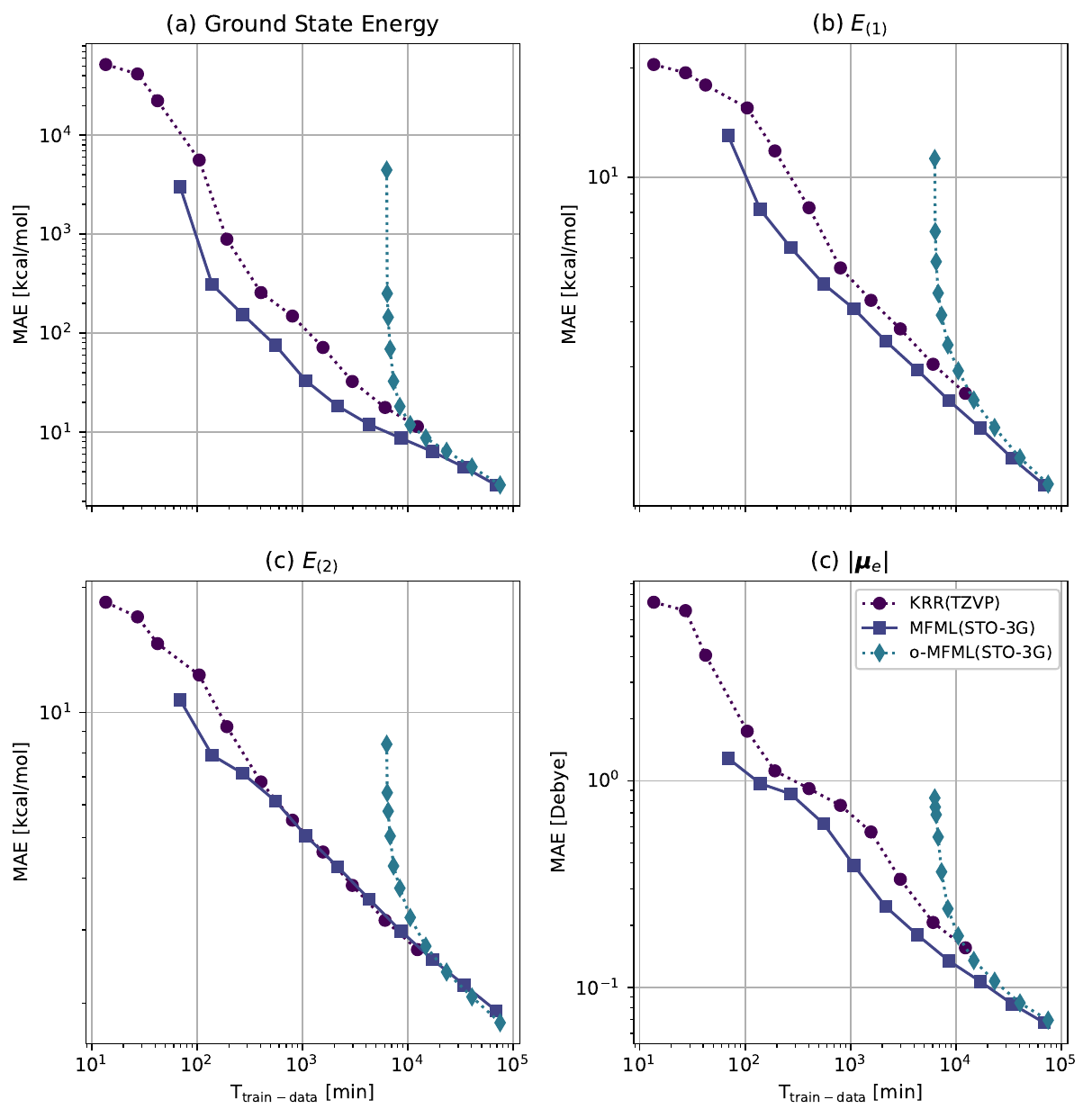}
    \caption{Time-cost versus MAE for MFML and o-MFML in comparison with single fidelity KRR.}
    \label{fig_time_MFML}
\end{figure*}

\begin{figure*}[htb!]
    \centering
    \includegraphics[width=0.8\textwidth]{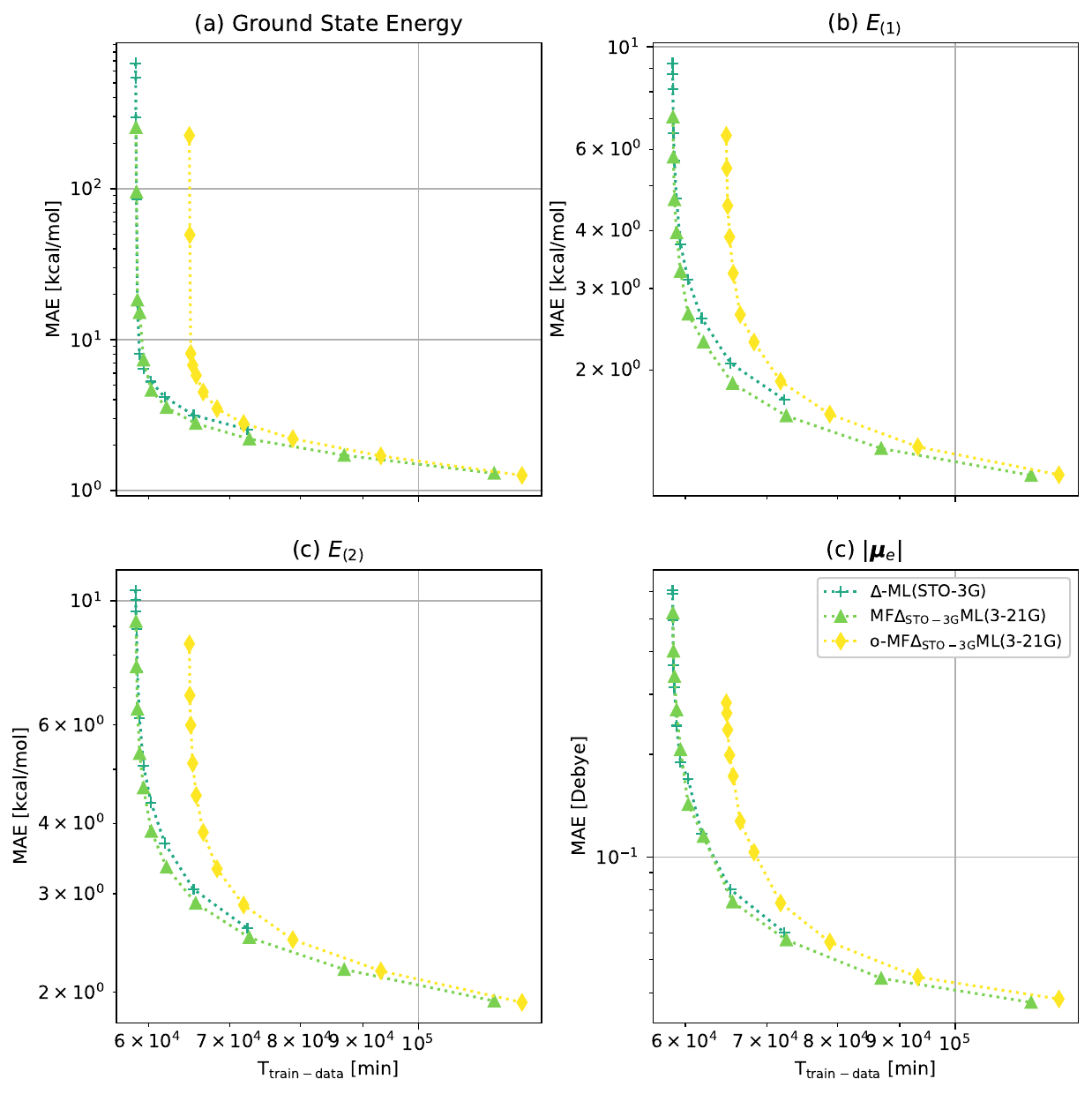}
    \caption{Time versus MAE for $\Delta$-ML, MF$\Delta$ML, and o-MF$\Delta$ML models in prediction of different QC properties.}
    \label{fig_time_delta}
\end{figure*}

The various ML methods studied in this work have promising results when the MAE is presented as a function of the training samples required at the target fidelity. However, to truly gauge the effectiveness of the ML models discussed above, the total cost to generate the training data is to be studied. 
For the MFML models, this corresponds to all the training data used in the multifidelity structure. 
This is computed as $T_{\rm MFML}=\sum_f N_f\cdot T_f$, where $T_f$ is the time to perform the QC computation for fidelity $f$ and $N_f$ is the number of samples used by the MFML model at this fidelity. For o-MFML this calculation would further include the cost of the validation set, $N_{val}\cdot T_F$. 
For $\Delta$-ML, the cost of the model for the prediction of some property for a molecule would be given as $T_{\Delta-ML} = N_{\rm train}^F\cdot(T_{f_b}+T_F) + T_{f_b}$. In cases several predictions are made, that is $N_{\rm test}$ number of predictions, then this cost becomes $T_{\Delta-ML} = N_F\cdot(T_{f_b}+T_F) + N_{\rm test}\cdot T_{f_b}$.
Note that $T_{F}$ and $T_{f_b}$ are different for different molecules from the QeMFi dataset. However, since the the dataset provides the compute costs for each molecule, this can readily be incorporated into the computation \cite{vinod2024QeMFi_paper, vinod_2024_QeMFi_zenodo_datatset}.
In short, for the $\Delta$-ML variants, the cost of the model includes the cost of making the calculations of the $QC_b$ for each geometry of the test set. Note that the term `test set' here is referred not to the small collection of molecules which would be used to assess the ML models before deployment. On the other hand, the large number of samples in this set allow it to be a meaningful proxy of actual use-case scenario where the ML model is used to predict the QC property of interest several times for several different geometries.
Note here, again, that for larger amounts of evaluations of the model, i.e.~for larger test sets, the total time for the required QC calculations grows as will be observed in the following parts of this manuscript.

The comparison of model MAE and time to generate training data for the diverse models studied in this work are shown in FIG.~\ref{fig_time_all}. The horizontal axis reports $T_{\rm train-data}$ which is the cost of generating training data for the ML models. For the $\Delta$-ML variants, this also includes the cost of the QC-baseline calculations.
FIG.~\ref{fig_time_all} displays the results for all the models studied in this work, namely,
\begin{enumerate}
    \item KRR with a single fidelity, in this case, TZVP,
    \item MFML and o-MFML with $f_b$: STO-3G,
    \item $\Delta$-ML with $QC_b$: STO-3G,
    \item MF$\Delta$ML and o-MF$\Delta$ML with $QC_b$: STO-3G and $f_b$: 3-21G.
\end{enumerate}
In the cases of the $\Delta$-ML variants, the QC-baseline is the QC method that is subtracted from all the fidelities which is not to be confused with $f_b$, which is the \textit{baseline fidelity} for the multifidelity methods. The MAE is reported in appropriate units of the QC property, while the time-cost is reported in minutes. While this figure serves the purpose of overall comparison of methods, each multifidelity method is compared separately in separate figures.

Consider the plot for the MFML curves seen in FIG.~\ref{fig_time_MFML}. 
For ground state energies as seen in FIG.~\ref{fig_time_MFML}(a), with around $10^{3}$ min time-cost, the MFML model results in MAE of $\sim30$ kcal/mol. The o-MFML model is shifted along the time-cost axis and results in a similar error for a cost of $\sim8\cdot10^3$ min. This is due to the the additional cost incurred to compute the validation set used in the optimization procedure involved in o-MFML \cite{vinod_2024_oMFML}. 
This offset is more pronounced in the Qc properties studied here since the o-MFML method did not additionally provide any improvement to the multifidelity model, possibly due to the conventional MFML combination already being optimal. In cases, where the o-MFML method does provide an improvement over conventional MFML method, such as that reported in ref.~\citenum{vinod_2024_oMFML}, o-MFML might result in a better MAE versus time-cost trade-off.
For example, consider FIG.~\ref{fig_time_MFML}(b) and FIG.~\ref{fig_time_MFML}(c) for the excitation energies. Here, although MFML initially shows better efficiency in terms of reaching a certain error for a lower time-cost that o-MFML, close to $5\cdot10^4$ min, the o-MFML method results in a lower, albeit marginally, MAE for the same time cost as compared to MFML. 

FIG.~\ref{fig_time_delta}(a)-(d) shows the time-cost versus MAE for the $\Delta$-ML, and MF$\Delta$ML variants for the prediction of diverse QC properties. These correspond to the furthest cluster of curves seen in FIG.~\ref{fig_time_all}. Due to the large test set size of 80,000 samples, the cost of the QC calculation of the baseline outweighs the plausible benefit of the $\Delta$-ML and its variants. Since fully trained ML models are generally used for a large number of predictions, it becomes evident that the use of $\Delta$-ML models for such cases becomes costly. Even so, the MF$\Delta$ML model performs better than the conventional $\Delta$-ML model, although not by a large margin. The cost of generating the QC baseline for the $\Delta$-ML models contributes the most to the overall cost of the models. The MFML (and by extension, o-MFML) method does not incur this cost since the baseline fidelity is predicted as opposed to QC computed. 
For each QC property, it is seen that the MF$\Delta$ML model is more efficient than the conventional $\Delta$-ML model. The optimized version, hence the o-MF$\Delta$ML model, is less efficient due to the additional cost of the validation set that is incurred. 

As one additional assessment, the time taken to merely train the models, that is given a training dataset, return a ML model that can be used for predictions, was also studied for the different ML models in this work. This is reported in the section S4 of the supplementary information associated with this manuscript. The curve of time taken to train the model as a function of the number of training samples used at the TZVP fidelity is shown in FIG.~S4. The explicit values of the time taken are reported in Table S1. For $N_{\rm train}^{\rm train}=2^{11}$, the time taken to train a single fidelity KRR model, and by extension, the standard $\Delta$-ML model was 0.8 seconds. For MFML with $f_b$:STO-3G, that is total of 5 fidelities, this time was $\sim 4300$ seconds, and for MF$\Delta$ML with $f_b:$3-21G, that is 4 total fidelities, this time was $\sim 440$ seconds.
These time-costs are marginal in comparison to the training data generation cost. Thus, this cost can be neglected as a contributing factor in the efficiency analysis of the multifidelity models.

\subsection{Large Test Set Sizes}
\begin{figure*}[htb!]
    \centering
    \includegraphics[width=0.8\textwidth]{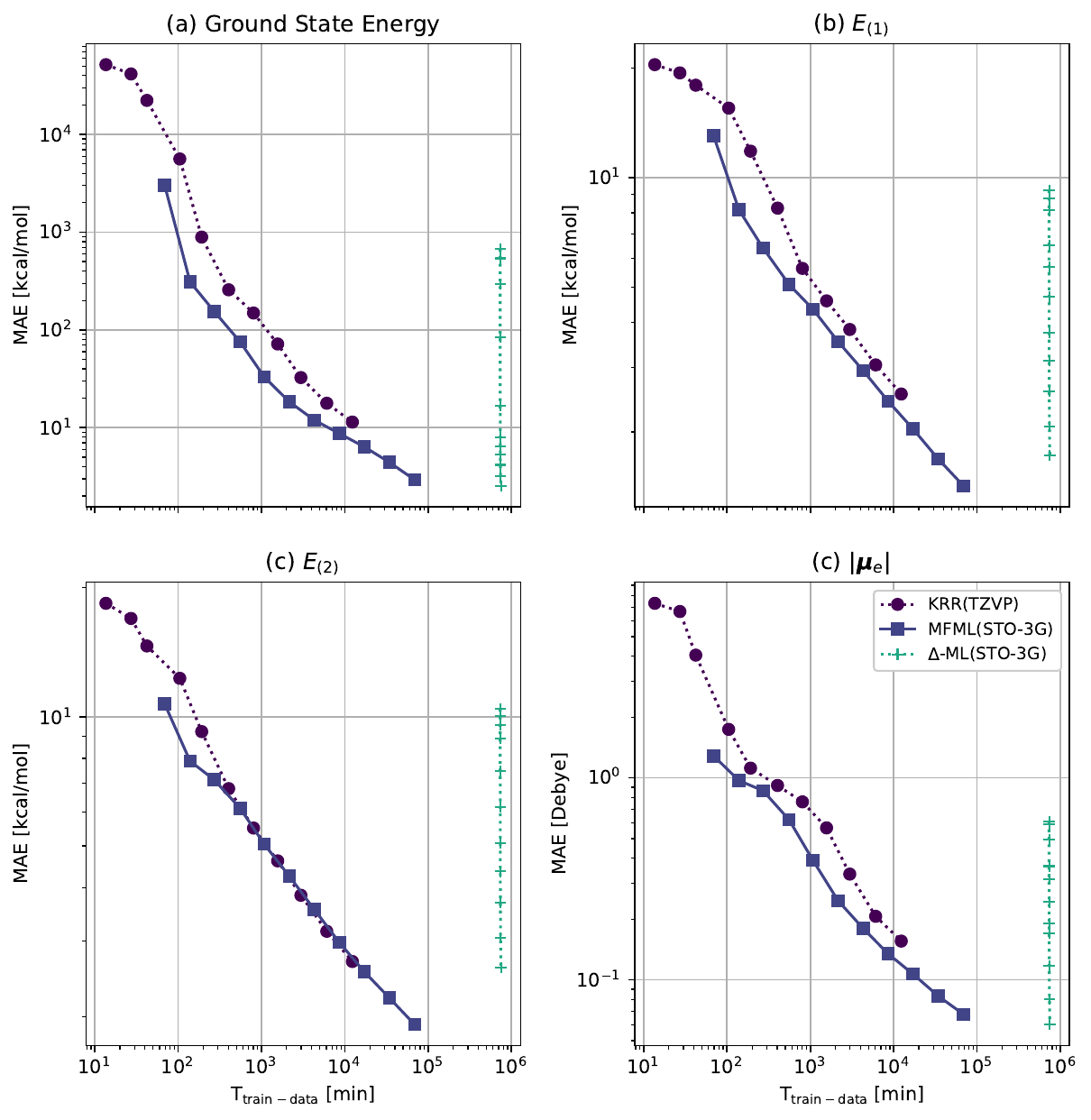}
    \caption{Time-cost versus model error for MFML and $\Delta$-ML for a hypothetical test set size of 1 million geometries.}
    \label{fig_time_hypothetical}
\end{figure*}

The above time-cost results reveal crucial aspects of using multifidelity methods for QC. When employing the ML-QC pipeline for large test set size, such as in the case of molecular trajectories, the MFML and o-MFML methods supersede the $\Delta$-ML variants. As the test set size increases, at the model accuracy versus cost trade-off for the $\Delta$-ML variants becomes difficult to justify. To really put in perspective the time-cost incurred in $\Delta$-ML, FIG.~\ref{fig_time_hypothetical} reports single fidelity KRR, MFML, and $\Delta$-ML MAE versus time-costs for the different QC properties for a hypothetical test set size of 1 million geometries. 
The single fidelity and MFML models are unaffected by the size of the test set and therefore do not show any change, the $\Delta$-ML model, however, shifts significantly to the right due to the cost incurred in the calculation of $QC_b$. The cost of using the $\Delta$-ML variants becomes a lot more evident here once again delineating the possible benefit of using the MFML method. 
As for the question of using a cheaper QC-baseline for $\Delta$-ML, the results of FIG.~\ref{fig_DeltaML_LC} and FIG.~S1 inform that the use of cheaper QC-baselines comes at the cost of model accuracy as is also evident from ref.~\citenum{Ramakrishnan2015}. The use of MFML is therefore recommended for cases where one is generally interested in predicting QC properties for a very large test set size. However, if the use-case only demands a small number of predictions, the $\Delta$-ML methods would still be a useful method. In particular, as shown in this work in FIG.~\ref{fig_deltaMFML_LC} and FIG.~\ref{fig_time_delta}, the MF$\Delta$ML method would then be preferred. 

\section{Conclusions and Outlook}
The work reported herein sets up time-cost benchmarks for $\Delta$-ML and MFML models that have been reported in the recent literature. In addition, the MF$\Delta$ML method was introduced and shown to be more effective than $\Delta$-ML. The MF$\Delta$ML method could be preferred over MFML in areas of application that require only a small number of model evaluations.
However, since such cases are rare in the field, and one generally predicts QC properties for a large collection of molecules or a long trajectory of a molecules, the MFML or o-MFML method are preferred as shown in this work.

Overall, this work has established
efficiency benchmarks for $\Delta$-ML and MFML methods along with their variants for four distinct QC properties of the QeMFi dataset. The use of MFML as a preferred method has been made clear with numerical evidence presented in the form of model error and the time-cost to achieve such error. In cases where $\Delta$-ML might be useful, the MF$\Delta$ML method introduced in this work is to be preferred for a low-cost high-accuracy model. The work presented here is a step in the direction of reducing the cost of training ML models for application in QC. Future research and applications in the ML-QC pipeline can use this work as reference to choose the most effective ML method to reduce the time-cost and can thereby present a time-cost benchmark that justifies the approach over those presented in this work. 
One possible extension of this current work is the assessment of the time-cost efficiency of $\Delta$-ML methods built on neural networks such as those discussed in ref.~\citenum{NN_deltaml_Liu22}. 

%%%end script
\begin{credits}
\subsection{Supplementary Material}
Supplementary sections S1-S4 and FIG.~S1-S4, Table S1.

\subsection{Data Availability}
The data used in this study from the QeMFi dataset can be found in \href{https://doi.org/10.5281/zenodo.13925688}{this ZENODO repository}. The programming scripts used for this benchmark can be openly accessed at \href{https://github.com/SM4DA/MFDeltaML}{this GitHub repository}.

\subsection{Acknowledgments}
The authors acknowledge support by the DFG through the project ZA 1175/3-1 as well as through the DFG Priority Program SPP 2363 on “Utilization and Development of Machine Learning for Molecular Applications – Molecular Machine Learning” through the project ZA 1175/4-1. The authors would also like to acknowledge the support of the `Interdisciplinary Center for Machine Learning and Data Analytics (IZMD)' at the University of Wuppertal.

\subsection{Disclosure of Interests}
There are no competing interests to declare.

\end{credits}

%%%%%%%%%% Merge with supplemental materials %%%%%%%%%%
\pagebreak
\begin{center}
\textbf{\large Supplementary Information: Benchmarking Data Efficiency in $\Delta$-ML and Multifidelity Models for Quantum Chemistry}
\end{center}
%%%%%%%%%% Merge with supplemental materials %%%%%%%%%%
%%%%%%%%%% Prefix a "S" to all equations, figures, tables and reset the counter %%%%%%%%%%
\setcounter{equation}{0}
\setcounter{section}{0}
\setcounter{figure}{0}
\setcounter{table}{0}
\setcounter{page}{1}
\makeatletter
\renewcommand{\theequation}{S\arabic{equation}}
\renewcommand{\thefigure}{S\arabic{figure}}
\renewcommand{\thepage}{S\arabic{page}} 
\renewcommand{\thesection}{S\arabic{section}}  
\renewcommand{\thetable}{S\arabic{table}}  
\renewcommand{\thefigure}{S\arabic{figure}} 
%%%%%%%%%% Prefix a "S" to all equations, figures, tables and reset the counter %%%%%%%%%%

\section{\texorpdfstring{$\Delta$}{Delta}-ML for QM7b}

The QM7b dataset consists of a total of 7,211 molecules with a maximum of seven heavy atoms \cite{montavon2013machine} . 
The atomization energies for each of these molecules is computed with 3 levels of Qc theory, namely, Hartree Fock (HF), Møller–Plesset perturbation theory (MP2) \cite{Quin_MP2_DFT_theory_2005, Yost_MP2_theory_2018, Pogrebetsky_MP2_theory_2023}, and Coupled Cluster Singles and Doubles perturbative Triples (CCSD(T)) \cite{Purvis_CCSD_theory_1982, Bartlett_CCSD_theory_2007, Crawford_CCSD_theory_2000}. For each level, 3 basis sets are used, STO-3G, 6-31G, and cc-pVDZ (with increasing size).
The atomization energy of a molecule is defined as the energy required to completely dissociate all the bonds of the molecule and break it into its constituent free atoms. Several works have evaluated the MFML and o-MFML method on this dataset \cite{zasp19a, vinod_2024_oMFML}. 
In order to assess whether the behavior of the $\Delta$-ML model as seen in the main text was influenced by the fidelities only varying by basis set choices, the same test was replicated for the QM7b dataset.

From the 7,211 geometries of the dataset, a random collection of 6,144 geometries were set aside as the training set, and the remaining were set aside as a test set. Since the QM7b dataset does not provide the compute times for the different fidelities, it is only used to check for the behavior of the $\Delta$-ML models across the different levels of QC theory, as opposed to varying basis set sizes. The data efficiency benchmarks are only made using the QeMFi dataset. Regardless, the QM7b dataset serves as a key indicator in studying the effects of varying QC theory levels as opposed to basis set sizes, which is the case for QeMFi. 

Based on previous research for the QM7b dataset in refs.~\cite{zasp19a, vinod_2024_oMFML}, the Laplacian kernel was used for KRR. This kernel is given as 
$$k\left(\boldsymbol{X}_i,\boldsymbol{X}_j\right) =\exp\left(-\frac{\lVert\boldsymbol{X}_i-\boldsymbol{X}_j\rVert_1}{\sigma}\right)~.$$
The kernel width parameter $\sigma$ was set to be 400.0, and the regularizer was set as $\lambda=10^{-10}$ for this assessment as prescribed in ref.~\cite{zasp19a}. 

\begin{figure}
    \centering
    \includegraphics[width=\linewidth]{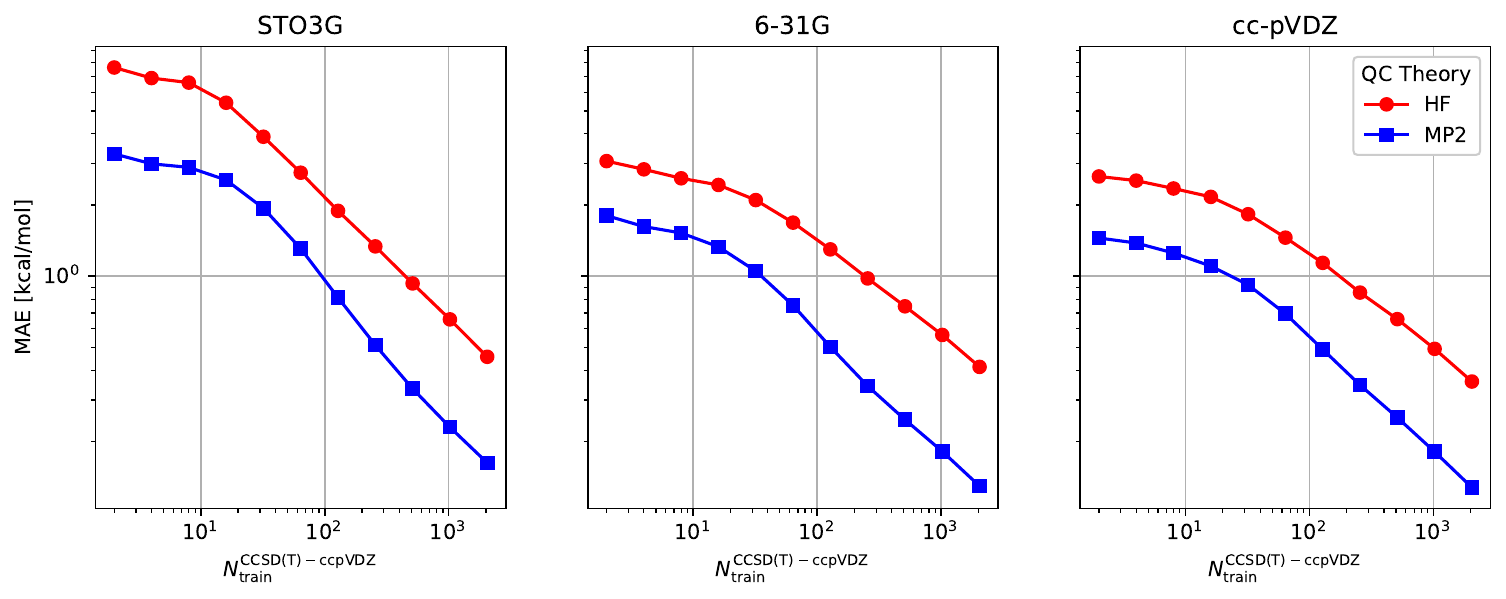}
    \caption{Learning curves for $\Delta$-ML with varying baseline fidelities for the atomization energies of the QM7b dataset. The different basis sets are denoted as subplot titles.}
    \label{fig_delML_QM7b}
\end{figure}

%In order to assess whether the results of the $\Delta$-ML method seen in the main text were merely due to the fidelities being described by basis set choice, with this dataset, a different approach was employed.
The basis set was fixed and the fidelities were then assumed to be the different QC levels of theory. This ensures that we only assess the effect of the level of theory and there are no artifacts arising from the basis sets.
That is, for each basis set that constitutes the multifidelity dataset of QM7b, the ordered fidelity in increasing order was considered to be HF, MP2, and then CCSD(T). Thus, for each basis set choice, the $\Delta$-ML model was built with $F=$CCSD(T) for HF and MP2 as $f_b$. The resulting learning curves are shown in Fig.~\ref{fig_delML_QM7b} for the different basis set choices.
The $\Delta$-ML model built with $f_b=$MP2 results in a lower model error in comparison to that built with HF as the baseline fidelity. This is observed regardless of the choice of the basis set.  
Thus it becomes evident that the results of the main text are not simply an artifact of the basis sets being set as fidelities. This is a general observation that the closer the baseline fidelity is to the target fidelity, the better the $\Delta$-ML model is at prediction of the QC property. 

\section{Predicting \texorpdfstring{$QC_b$}{QC\_b} for \texorpdfstring{$\Delta$}{Delta}-ML}
\begin{figure*}[htb!]
    \centering
    \includegraphics[width=\linewidth]{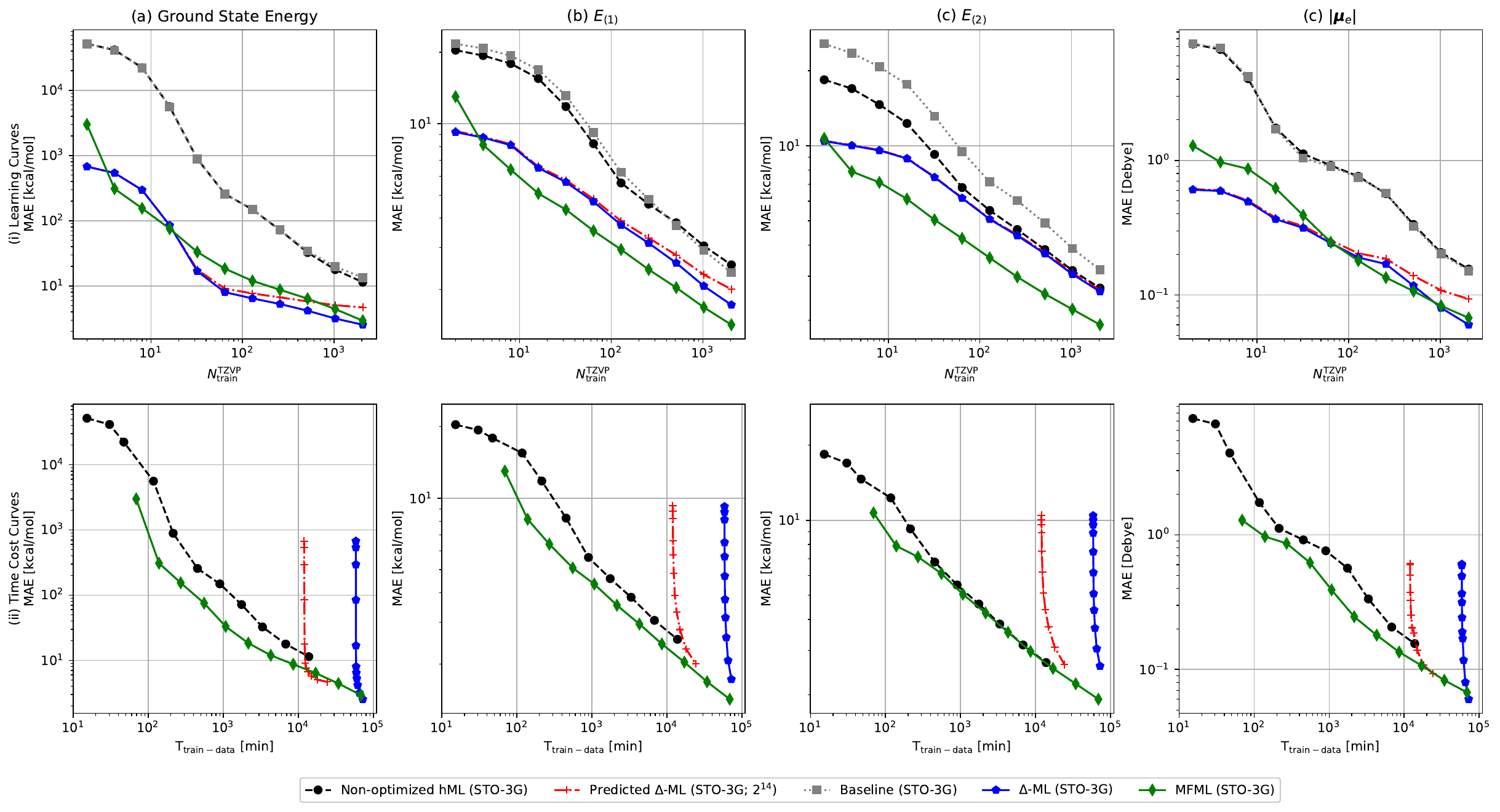}
    \caption{
    Learning curves (row (i)) and time-cost assessment (row (ii)) for a non-optimized two fidelity hML model, a predicted $QC_b$ $\Delta$-ML variant, and MFML model for the prediction of ground state energies, first and second vertical excitation energies, and magnitude of electronic contribution to molecular dipole moments. The error in prediction of $QC_b$ with the non-optimized hML model is also depicted (in gray). The predicted $QC_b$ model is trained on $2^{14}$ training samples at the STO-3G fidelity.
    }
    \label{fig_predicted_Delta}
\end{figure*}

Based on the results for $\Delta$-ML variants and MFML variants from the main text, a natural line of research is to consider a case for $\Delta$-ML, where one does not need to perform the baseline QC computations but uses an ML model to predict those QC-baseline energies instead. This would be somewhat similar to what is performed in MFML for $f_b$. Two such models are studied in this section.
First, since the $\Delta$-ML model uses the baseline fidelity to train the differences, these same are used to train a model for the QC-baseline prediction over the test set. If the $\Delta$-ML model uses 2 samples, then these same samples are used to predict the baseline over the test set. Thus there is no additional cost incurred in the this \textit{predicted}-$\Delta$-ML model. This model is equivalent to a 2-level hML \cite{dral2020hierarchical} model built with the same number of training samples at both fidelities, that is, a hML model that does not run the \textit{ad hoc} optimization procedure for the number of training samples. Therefore, this model is referred to hereon as non-optimized hML model.
Second, a model is trained with $2^{14}$ samples for the STO-3G fidelity to predict $QC_b$. The predictions from this model are used to replace $y^{QC_b}$ from Eq.~\eqref{delML_pred}. This model is referred to as `Predicted $\Delta$-ML (STO-3G; $2^{14}$)'. 

The analysis for the predicted-$\Delta$-ML model is shown in FIG.~\ref{fig_predicted_Delta} along with the conventional $\Delta$-ML and MFML methods for the prediction of the different QC properties studied in this work. 
Row (i) shows the learning curves as a function of the number of training samples used at TZVP. In addition, the learning curve for the prediction of $QC_b$ used in the non-optimized hML model over the test set is provided for reference. 
The results indicate that the error of predicting the baseline QC-fidelity is a huge contributor to the overall error of the predicted-$\Delta$-ML model, as also stated in ref.~\citenum{dral2020hierarchical}, where the authors report that the use of identical number of training points for all fidelities does not provide any benefit over the single fidelity model in terms of model error. 
The two curves are almost entirely overlaid on each other for the most part with only a small deviation observed for $N_{\rm train}^{\rm TZVP}=2^{11}$ for most of the QC properties. For $E_{(2)}$, the error of the prediction of the QC-baseline is higher possibly due to the added chemical complexity of the second vertical excitation state. Although the learning curves for $\Delta$-ML and MFML were already discussed in FIG.~\ref{fig_DeltaML_LC} and FIG.~\ref{fig_MFML_LC} respectively, here they are visible in contrast with each other. The two learning curves seem to converge to similar MAE values for large training set sizes. Further, the Predicted $\Delta$-ML (STO-3G; $2^{14}$) model initially has MAE comparable to the $\Delta$-ML model for the case of ground state energies but saturates for $N_{\rm train}^{\rm TZVP}>2^7$. For $E_{(1)}$, $E_{(2)}$, and $\lvert\boldsymbol{\mu}_e\rvert$ this model has MAE comparable to the MFML model for the most part with some saturation observed for larger training set sizes. 

FIG.~\ref{fig_predicted_Delta}(ii) studies the time-cost analysis of the two predicted $QC_b$ models in contrast with the $\Delta$-ML and MFML models. 
First it is observed that the non-optimized hML model does not provide any benefit over MFML as was reasoned above. Second, the predicted $\Delta$-ML (STO-3G; $2^{14}$) model for all QC properties is shifted to the right-hand side of the plots due to the cost of training data for the prediction of $QC_b:$ STO-3G. For all QC properties, the MFML method has a lower MAE than this predicted $\Delta$-ML variant for a given time-cost. 
The $\Delta$-ML model is shifted by a large time-cost that is incurred due to the QC-baseline calculation as was discussed previously in light of FIG.~\ref{fig_time_delta}. These results from FIG.~\ref{fig_predicted_Delta} strongly indicate that the use of the predicted $\Delta$-ML variants does not provide any foreseeable benefit.

\section{Validation Set and o-MFML Learning Curves}
\begin{figure}[htb!]
    \centering
    \includegraphics[width=0.5\linewidth]{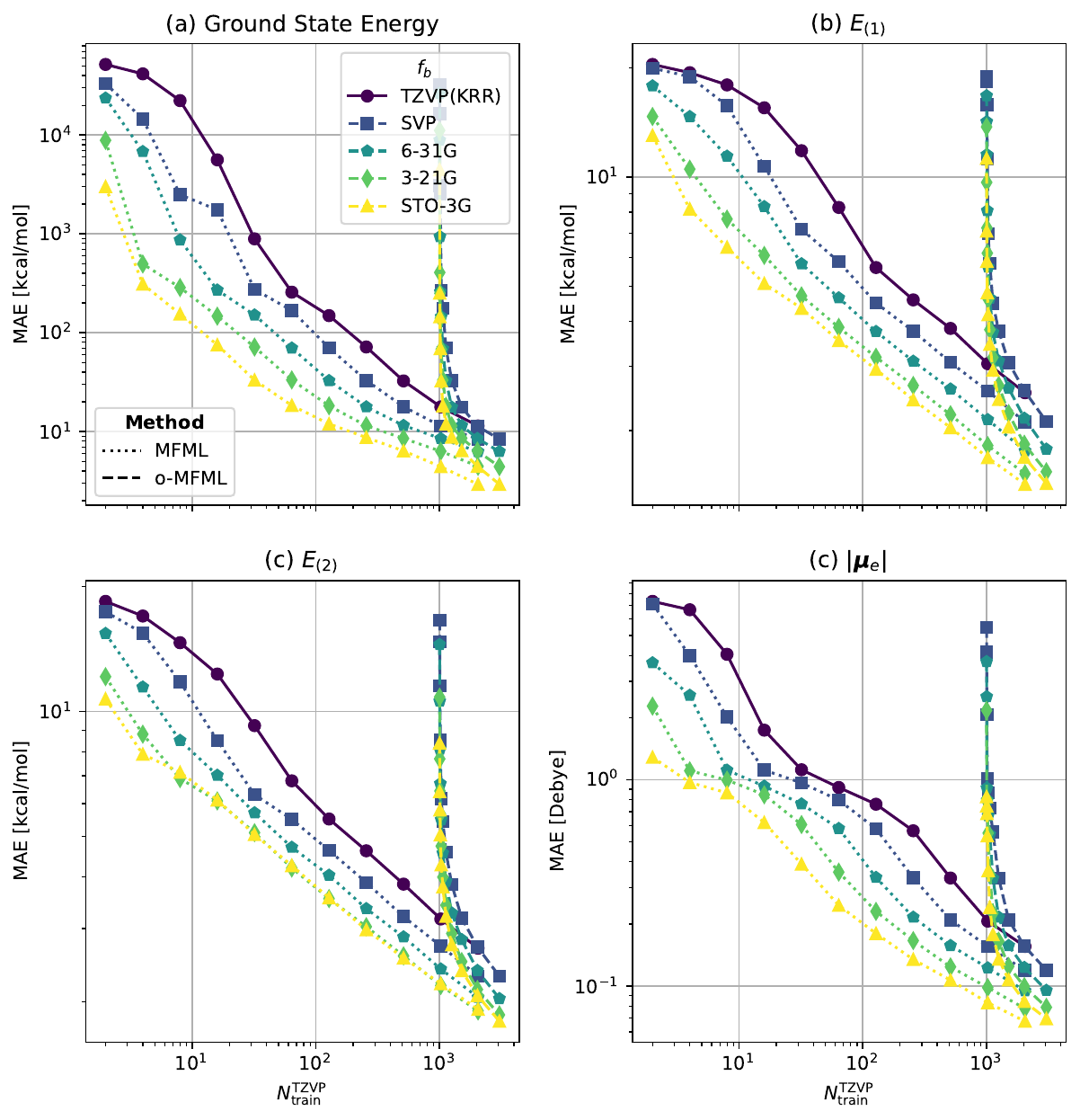}
    \caption{MFML and o-MFML learning curves for the various QC properties with the fvalidation set size being accounted for o-MFML.}
    \label{fig_valsetsize_omfml_LC}
\end{figure}

\section{Training Time of the Models}

\begin{figure}[htb!]
    \centering
    \includegraphics[width=0.5\linewidth]{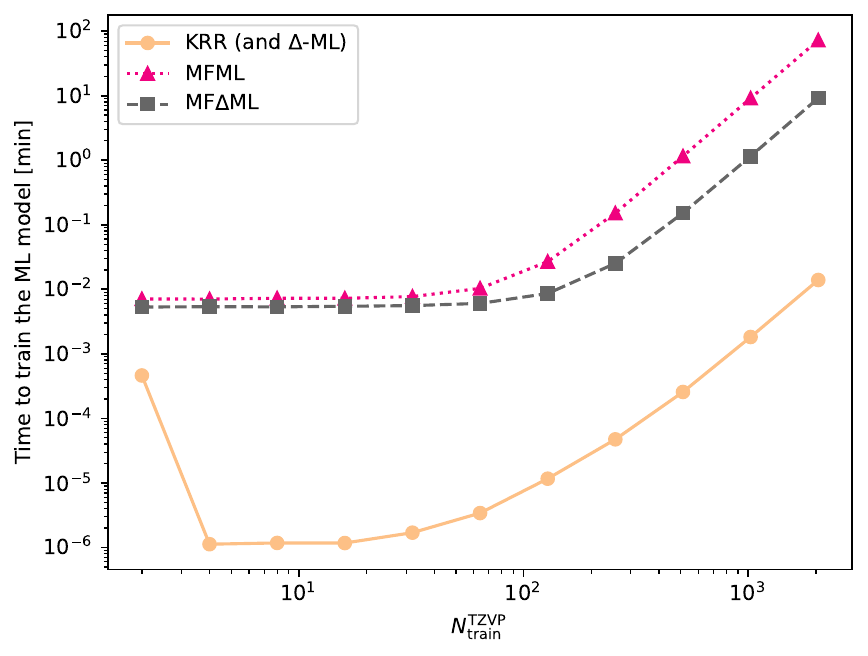}
    \caption{Time to train different ML models as a function of number of training samples used at the TZVP fidelity. The time is reported here in minutes to provide ready comparison to the time-cost assessment curves of the main text.}
    \label{fig_train_time_model}
\end{figure}

\begin{table}[htb!]
    \centering
    \begin{tabular}{|c|c|c|c|}
        \hline
        $\boldsymbol{N_{\rm train}^{\rm TZVP}}$ & KRR & MFML & MF$\Delta$ML\\
        \hline
        \hline
        2 & 0.0277 & 0.4257 & 0.3188 \\
        4 & 0.0001 & 0.4232 & 0.3226 \\
        8 & 0.0001 & 0.4364 & 0.3209 \\
        16 & 0.0001 & 0.4353 & 0.3277 \\
        32 & 0.0001 & 0.4631 & 0.3354 \\
        64 & 0.0002 & 0.6198 & 0.3632 \\
        128 & 0.0007 & 1.6126 & 0.5163 \\
        256 & 0.0028 & 9.1021 & 1.5134 \\
        512 & 0.0154 & 69.3421 & 9.1189 \\
        1024 & 0.1095 & 548.575 & 69.2028 \\
        2048 & 0.8366 & 4343.7689 & 550.4958 \\
        \hline
    \end{tabular}
    \caption{Time taken to train different ML models studied in this work based on number of training samples chosen at the TZVP fidelity. The time is reported here in seconds and rounded to the fourth decimal.}
    \label{tab_traintimes}
\end{table}

\bibliography{main}

\end{document}